\documentclass[superscriptaddress,aps,pra,twocolumn,showpacs,preprintnumbers,amsmath,amssymb,floatfix,groupedaddress]{revtex4}

\def\e{\mathcal{E}}
\def\ds{\delta_s}
\def\dt{\widetilde{\delta}}
\newcommand{\ket}[1]{|{\rm #1}\rangle}
\newcommand{\bra}[1]{\langle {\rm #1}|}
\usepackage{graphicx,bm}
\begin{document}


\markboth{N.\ B.\ Phillips, A.\ V.\ Gorshkov, and I.\ Novikova}{Slow light propagation and amplification via
electromagnetically induced transparency and \dots}


\title{Slow light propagation and amplification via electromagnetically induced transparency and four-wave mixing in an optically dense
atomic vapor}

\author{Nathaniel B. Phillips}
\affiliation{Department of Physics, College of William and Mary, Williamsburg,
Virginia 23185, USA}
\author{Alexey V. Gorshkov}
\affiliation{Department of Physics, Harvard University, Cambridge,
Massachusetts 02138, USA}
\author{Irina Novikova}
\affiliation{Department of Physics, College of William and Mary, Williamsburg,
Virginia 23185, USA}

\date{\today}

\begin{abstract}
We experimentally and theoretically analyze the propagation of  weak signal field pulses under the conditions of
electromagnetically induced transparency (EIT) in hot Rb vapor, and study the effects of resonant
four-wave mixing (FWM). In particular, we demonstrate that in a double-$\Lambda$ system, formed by the strong
control field with the weak resonant signal and a far-detuned Stokes field, both continuous-wave spectra and pulse
propagation dynamics for the signal field depend strongly on the amplitude of the seeded Stokes field, and the
effect is enhanced in optically dense atomic medium. We also show that the theory describing the coupled
propagation of the signal and Stokes fields is in good agreement with the experimental observations.
%
\end{abstract}

\pacs{42.50.Gy, 32.70.Jz, 42.50.Md}

\maketitle

\section{Introduction}

In recent years, electromagnetically induced transparency (EIT)~\cite{lukin03rmp,marangosEITreview} and
the associated effect of ultra-slow pulse propagation (``slow light'')~\cite{boydPiO02} have attracted considerable
attention due to their many promising applications. For example, they enable coherent, reversible transfer
between the quantum states of an electromagnetic field and the collective excitation of an ensemble of long-lived
radiators (e.g., spins of atoms, solid-state impurities, quantum nanostructures, etc.), which are necessary for the
realization of quantum memory~\cite{lukinPRL00}, entanglement of distant matter nodes \cite{DLCZ},
single photon generation~\cite{simonPRL2007}, and realization of deterministic two-qubit gates for photons~\cite{muschikPRL08}, etc. At the same time, more robust control of the propagation of classical pulses using
EIT has been extensively explored in various materials for applications such as optical packet switching and optical signal processing~\cite{TuckerJLT}.

In a traditional EIT scheme, a strong classical control field is applied to one optical transition, resulting in
a modification of the optical properties of a weak signal field, which couples the same excited electronic state
with a second long-lived ground-state sublevel, thereby forming a $\Lambda$ system~\cite{lukin03rmp}, as
depicted in Fig. \ref{figure1}(a).  In the limit of low optical depth, it is sufficient to take into account only
the effects of this single $\Lambda$. However, many applications require operation at high optical
depth~\cite{boydPRA05,gorshkovPRL07}, where additional nonlinear effects may become
important~\cite{kangPRA04,wongPRA04,haradaPRA06,agarwalPRA06}. One such effect is resonant four-wave mixing---a
nonlinear process arising from the far off-resonant interaction of the control field. Earlier
studies~\cite{lukinPRL98,lukinPRL99} found that the propagation of the signal field in this case will be
strongly affected by the presence of the Stokes field.

The effect of four-wave mixing can be advantageous or detrimental, depending on the details of the application.
For example, in quantum memory applications the resonant mixing reduces the fidelity by adding extra noise into the signal field. Also, FWM may limit the storage efficiency at higher optical depth
~\cite{phillipsPRA08}. On the other hand, non-classical correlations between two signal and Stokes fields can
individually carry quantum information ~\cite{lett} and produce entangled images. Similarly, for slow light
applications, the conversion of an original pulse from the signal to Stokes channel may reduce the readout
efficiency ~\cite{matskoOL05}. However, under certain conditions, FWM may lead to gain for both the signal and
Stokes fields, which could compensate for any optical losses~\cite{camachoNature}.

In this manuscript, we investigate the modification of the two-photon resonant transmission peak and the
dynamics of pulse propagation in the case of a seeded Stokes field, which was produced simultaneously with
the signal field by phase modulating the original monochromatic control field.  Specifically, we show that the
resulting signal and Stokes spectra are well-described by a double-$\Lambda$ system.  Additionally, we show that
this model accurately portrays the dynamics of signal and Stokes pulse propagation.

\section{Experimental Arrangements}
\begin{figure}
\begin{center}
\includegraphics[width=1\columnwidth]{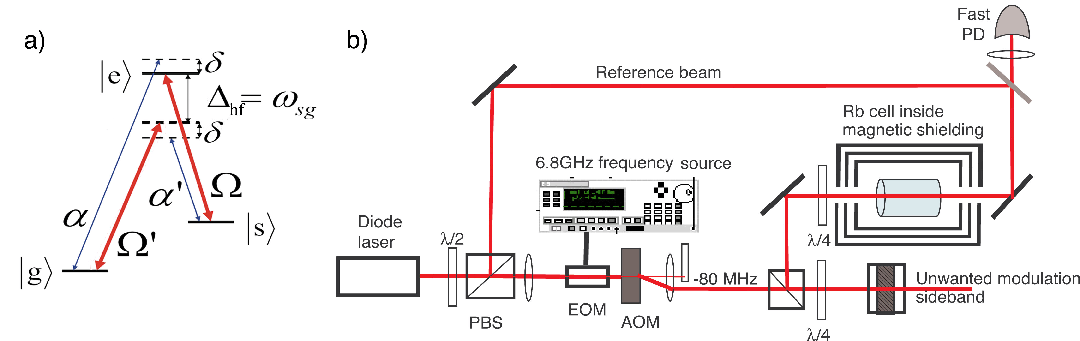}%
\caption{(Color Online)  (a)  The double-$\Lambda$ system used in theoretical calculations.  (b)  A schematic of
the experimental apparatus (see text for abbreviations).} \label{figure1}
\end{center}
\end{figure}
Measurements were performed using the configuration in Fig.~\ref{figure1}(b).  We tuned a commercial external
cavity diode laser (ECDL) near the Rubidium $D_1$ transition ($\lambda = 795$ nm).  After using a polarizing
beam splitter (PBS) to separate a fraction of the light for a reference beam, we passed the main beam through an
electro-optical modulator (EOM), which modulated its phase at the frequency of the ground state hyperfine
splitting of $^{87}$Rb ($\Delta_{\rm hf}/2 \pi = 6.835$ GHz). Due to the small driving amplitude, the phase
modulation mostly produced two first modulation sidebands at $\pm \Delta_{\rm hf}$ of nearly equal amplitudes
and opposite phases. We tuned the zeroth order (carrier frequency) field to the $5^2\rm{S}_{1/2}{\rm F}=2 \to
5^2\rm{P}_{1/2}{\rm F'}=2$ transition; this beam acted as the control field. The $+1$ modulation sideband of the
amplitude functioned as the signal field, and was tuned near the $5^2\rm{S}_{1/2}{\rm F}=1 \to
5^2\rm{P}_{1/2}{\rm F}=2$ transition. The $-1$ sideband acted as the far-detuned Stokes field. Then all optical
fields passed through an acousto-optical modulator (AOM) operating at 80 MHz, which shifted the frequencies of
the fields by that amount.

For spectral measurements, the control field was always on, thereby ensuring that most of the atoms were in
$\ket{g}$, and we slowly swept modulation frequency of the EOM, which synchronously scanned the two-photon
detuning for the Stokes and signal fields. For slow light measurements, we first applied a pulse of the control
field to optically pump the atoms into $\ket{g}$, and then we adjusted the modulation power of the EOM and AOM
to produce desired Gaussian pulses of the signal and Stokes fields.

To carefully evaluate the effects of resonant four-wave mixing, we used  a temperature tunable Fabry-P\'erot
etalon (FSR = 20 GHz, finesse $\approx$ 100) to reduce the Stokes field amplitude by tuning the etalon
transmission resonance such that most of the $-1$ sideband is transmitted, but all the other fields are
reflected. This way, we were able to reduce the intensity of the Stokes field by a factor of $20$ without
noticeable losses in the intensity of both the control and signal fields.

Before entering a vapor cell, the beam was weakly focused to either 2.6 mm or 4 mm diameter, as we indicate
below, and circularly polarized with a quarter-wave plate ($\lambda/4$).  Typical peak control field and signal
field powers were approximately 19 mW and 50 $\mu$W, respectively.  A cylindrical Pyrex cell, of length 75 mm
and diameter 22 mm, contained isotopically enriched $^{87}$Rb and 30 Torr Ne buffer gas, so that the
pressure-broadened optical transition linewidth was $2\gamma = 2\pi \times 290$ MHz~\cite{rotondaro}.  The cell
was mounted inside tri-layer magnetic shielding, as to reduce the effects of stray magnetic fields. The
temperature of the cell (and thus the concentration of Rb in the vapor phase) was adjusted using a bifilar
resistive heater wound around the innermost magnetic shielding layer in the range between $70^{\circ}$C and
$80^{\circ}$C, which corresponded to the change in Rb densities from $5.6\times10^{11}$~cm$^{-3}$ and
$1.2\times10^{12}$~cm$^{-3}$, and to the range of optical depths $2d$ between 52 and 110. Here we define the
optical depth $2d$ such that the probe intensity without EIT is attenuated by $\mathrm{e}^{-2d}$, and our
procedure for calculating the effective optical depth 
is described in
Ref.~\cite{phillipsPRA08}.  We also measured~\cite{phillipsPRA08} the typical spin wave decay time to be $1/(2 \gamma_{sg}) \simeq 500~\mu$s,
most likely arising from small, uncompensated remnant magnetic fields.

After the cell, the output laser fields were recombined with the unshifted reference beam at a fast
photodetector, and the beat note signals between each of the $+1$ and $-1$ modulation sidebands and the
reference field was measured using a microwave spectrum analyzer. Because of the 80~MHz frequency shift
introduced by the AOM, the different beat note frequencies of each sideband with the reference field allowed for independent measurement of their amplitudes.

\section{Theoretical model}
In a traditional three-level system, under the conditions of electromagnetically induced transparency, a strong
control field (frequency $\nu_{\rm C} = \omega_{es}$, Rabi frequency $\Omega$)~\cite{rabicomment} works in
conjunction with a weak signal field ($\nu= \omega_{eg}+\delta$, Rabi frequency $\alpha$) to create a long-lived
coherence between states $\ket{g}$ and $\ket{s}$ \cite{lukin03rmp}, as depicted in Fig.~\ref{figure1}(a),
producing a narrow symmetric transparency peak in the spectrum of the signal field near the two-photon resonance
($\nu-\nu_{\rm C} \approx \omega_{gs}$). Simultaneously, the signal field experiences  a steep variation in refractive index, thereby reducing the pulse's group velocity (``slow
light''), and leading to a pulse delay time of $\tau = (d \gamma)/|\Omega|^2 \gg L/c$~\cite{boydPiO02}, where $\gamma$ is the optical polarization decay rate.

While this description is sufficient in the limit of low optical depth~\cite{phillipsPRA08}, at high optical
depth, accompanying nonlinear processes become important. A sufficiently strong control field can excite the far off-resontant
transition from $\ket{g}$ to $\ket{e}$ \cite{lukinPRL98}, and generate a Stokes field $\alpha'$ via the
four-wave mixing (FWM) process.
We theoretically model this effect by considering a double-$\Lambda$ configuration, consisting of a near-resonant
$\Lambda$ system formed by the control and signal fields and of the additional far-detunied $\Lambda$ system
formed by the same control field ($\nu_{\rm C} = \omega_{es}$, Rabi frequency $\Omega'$), applied to the state
$\ket{g}$ and by an additional Stokes field ($\nu' = \omega_{es} - \Delta_{\rm hf} - \delta$, Rabi frequency
$\alpha'$).
In such system, we can  use Floquet theory \cite{dreseEPJD99} to adiabatically eliminate the off-resonant
interaction via $\Omega'$ and $\alpha'$. As a result, to first order in $1/\Delta_{\rm hf}$ and in $\alpha'$,
one obtains an effective Rabi frequency $\Omega' \alpha'^*/\Delta_{\rm hf}$ coupling $\ket{g}$ and $\ket{s}$,
while the states $\ket{e}$ and $\ket{g}$ acquire small light shifts $\delta_s = |\Omega'|^2/\Delta_{\rm hf}$ and
$-\delta_s$, respectively. For our Clebsch-Gordan coefficients \cite{phillipsPRA08}, $|\Omega'|^2 = 3
|\Omega|^2$.

The rotating-frame Hamiltonian describing this interaction is:
%
%
\begin{equation}\begin{split}
\hat{H}=& -(\delta -\ds) \ket{s}\bra{s} - (\delta- 2 \ds)\ket{e}\bra{e} \\
&- \left[ \alpha \ket{e}\bra{g} + \Omega
\ket{e} \bra{s} + \frac{\Omega' \alpha'^*}{\Delta_{\rm hf}}\ket{s}\bra{g} + {\rm H.c.} \right].
\end{split}
\end{equation}
Here, the Rabi frequencies of the signal and the Stokes fields are $\alpha = \e \mu/\hbar$ and $\alpha' = \e'
\mu'/\hbar$, where $\e$ and $\e'$ are the corresponding slowly-varying envelopes, and $\mu$ and
$\mu'$ are the (real) dipole matrix elements of the respective transitions.

In the undepleted pump and adiabatic approximations, the Fourier components of the signal field $\e(\omega)$ and
the Stokes field $\e'(\omega)$ propagate according to the coupled differential equations, to linear order in $\alpha$ and $\alpha'$ \cite{lukinPRL97,lukinPRL98, lukinPRL99}:%
\begin{widetext}
\begin{equation}
\partial_z \left[ \begin{array}{c}
\e(z,\omega) \\
\e'^*(z,\omega)
\end{array}\right] = i \frac{d \gamma}{F L}
\left[\begin{array}{cc}
\delta-\ds + \omega + i \gamma_0 & -\Omega^2/\Delta_{\rm hf} \\
\Omega^{*2}/\Delta_{\rm hf} & 0 \end{array} \right] \left[ \begin{array}{c}
\e(z,\omega) \\
\e'^*(z,\omega) \end{array} \right], \label{diffeq}
\end{equation}
\end{widetext}%
where we take into account the optical polarization decay rate $\gamma$ and the ground state decay rate
$\gamma_0\approx \gamma_{sg}+ \gamma \frac{|\Omega'|^2}{\Delta_{\rm hf}^2}$ \cite{axelthesis}, and have set
$F= |\Omega|^2 + [\gamma  - i (\delta-2\ds+\omega)][\gamma_0 - i (\delta-\ds+\omega)]$.
Under the assumption that $\Omega$ is real and z-independent, Eq.~\eqref{diffeq} can be solved analytically for
$\e(z, \omega)$ and $\e'^*(z,\omega)$, for the conditions corresponding to our experiment: $\e'^*(0,\omega) = -f
\e(0,\omega)$: the Stokes seed has the same initial temporal lineshape as the signal pulse, but with an opposite
phase and with some amplitude scaling factor $0 < f \le 1$. Defining the Raman detuning as $\Delta_{\rm R} =
-\Omega^2/\Delta_{\rm hf}$, and with $\beta(\omega) \equiv \sqrt{[\gamma_0-i(\delta-\ds+\omega)]^2+ 4\Delta_{\rm
R}^2}$, $\sigma(\omega) \equiv \frac{1}{2}\frac{d \gamma}{F L}(\delta -\ds + \omega + i \gamma_0)$, and
$\xi(\omega) \equiv \frac{1}{2}\frac{d \gamma}{F L}\beta(\omega)$, we find the following analytic expressions
for the Fourier components of the signal and Stokes fields \cite{lukinPRL99, hongPRA09}:
\begin{widetext}
\begin{eqnarray}
\e(z,\omega) &=& \e(0,\omega){\rm e}^{i \sigma(\omega) z}\left[\cosh[\xi(\omega) z] + i \left( \frac{\sigma(\omega)}{\xi(\omega)} - f  \frac{2 \Delta_{\rm R}}{\beta(\omega)}\right) \sinh[\xi(\omega)z] \right], \label{signalexpr}\\
\e'^*(z,\omega) &=& -f\e(0,\omega) {\rm e}^{i \sigma(\omega) z}\left[\cosh[\xi(\omega) z] - i
\left(\frac{\sigma(\omega)}{\xi(\omega)} -  \frac{1}{f}\frac{2 \Delta_{\rm R}}{\beta(\omega)}\right)
\sinh[\xi(\omega)z] \right]. \label{stokesexpr}
\end{eqnarray}
\end{widetext}
Eqs.~\eqref{signalexpr} and \eqref{stokesexpr} fully describe the propagation of the light fields through the
atomic medium.  Theoretically, the measured transmission spectra of the signal and Stokes fields are computed as
$|\e(L)|$ and $|\e'(L)|$, respectively using Eqs.~\eqref{signalexpr} and \eqref{stokesexpr}.  We interpret equations \eqref{signalexpr} and
\eqref{stokesexpr} by first applying a few simplifications, similar to Ref.~\cite{hongPRA09}, by first shifting the two-photon detuning by the light shift, defining $\dt=\delta-\ds$ (setting
$\omega=0$ for continuous wave measurements), and considering large $|\dt| \gg2 |\Delta_{\rm R}|$, but also $|\dt| \ll\Omega$, and $|\dt| \gg\gamma_0$.  Under these assumptions, $\beta \approx i \dt$, and for our parameters, $2\xi L = 2 i \sigma L \approx
i \frac{\dt}{\Omega^2/(d \gamma)} -\frac{\dt^2}{[\Omega^2/(\sqrt{d}\gamma)]^2}$, where the denominators of the two terms represent, respectively, the inverse of the EIT group delay and the square of the width of the EIT transparency window.  Under these approximations, which hold well in our experiments, the signal and Stokes
amplitudes after the cell are:
\begin{eqnarray}
|\e(L)|&=&\e(0){\bigg |}e^{2i\sigma L}-f\frac{\Omega^2}{\Delta_{\rm hf}\dt}(1-e^{2i\sigma L}){\bigg |}, \label{probeapprox}\\
|\e'(L)|&=&\e(0){\bigg |}\frac{\Omega^2}{\Delta_{\rm hf}\dt}(1-e^{2i\sigma L})-f{\bigg |}.
\end{eqnarray}

These expressions allow us to interpret the transmission spectra for both fields in terms of an interference
between  EIT  and FWM effects, where the FWM effect is represented by the terms proportional to $1/\Delta_{\rm
hf}$~\cite{hongPRA09}. At small optical depth, the effect of FWM on the signal field transmission is negligible,
and we observe a typical symmetric EIT transmission peak (not shown). As optical depth increases,
the FWM term ($\propto 1/\Delta_{\rm hf}$) in Eq.~(\ref{probeapprox}) becomes more noticeable relative to the
EIT term, which reduces with $d$ due to the narrowing of the EIT window.
Since the phase of $e^{2 i \sigma L}$ in Eq.~(\ref{probeapprox}) is $\approx \dt/(v_g/L)$,
for $\dt>0$, $\dt = n\pi v_g/L$ gives destructive interference (and hence dips in the spectrum) for even $n$,
and constructive interference for odd $n$.  For $\dt<0$, the opposite case is true: even $n$ yields constructive
interference; odd $n$ yields destructive interference.

\section{Spectral Measurements}
\begin{figure*}[!htdp]
\begin{center}
\includegraphics[width=1.85\columnwidth]{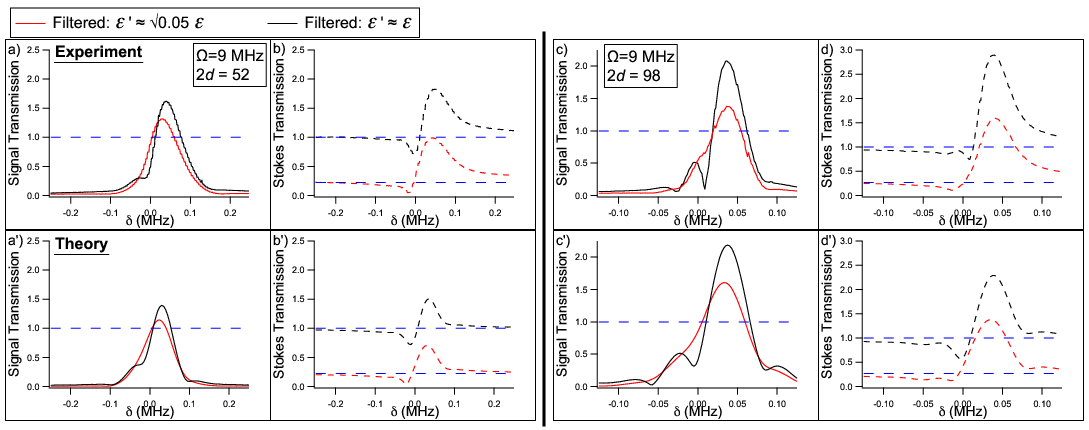}%
\caption{(Color Online)  (a, b) Signal and stokes amplitude spectra at an optical depth of $2d=52$ with a beam diameter of 4 mm and $\Omega/2\pi = 9$ MHz.  Black traces are with a full Stokes seed present.  Red traces are with the Stokes seed amplitude attenuated to $\approx \sqrt{0.05}$ of the signal field's amplitude.  (a$^\prime$, b$^\prime$) Corresponding theoretical predictions.  (c, d)  Signal and stokes amplitude spectra at an optical depth of $2d=98$.  (c$^\prime$, d$^\prime$) Corresponding theoretical predictions.}%
\label{figure2}
\end{center}
\end{figure*}
\begin{figure*}[h!tdp]
\begin{center}
\includegraphics[width=1.85\columnwidth]{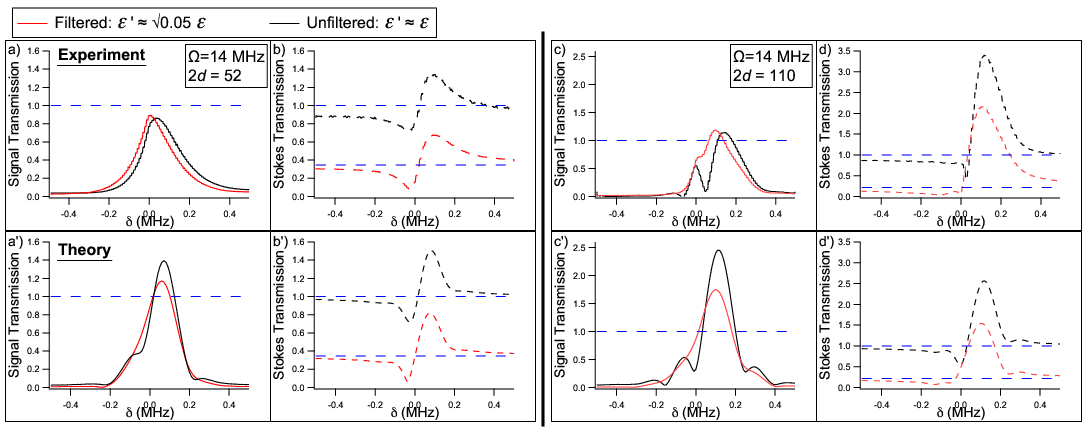}%
\caption{(Color Online)  (a, b) Signal and stokes amplitude spectra at an optical depth of $2d=52$ with a beam diameter of 2.6 mm and $\Omega/2\pi = 14$ MHz.  Black traces are with a full Stokes seed present.  Red traces are with the Stokes seed amplitude attenuated to $\approx \sqrt{0.05}$ of the signal field's amplitude.  (a$'$, b$'$) Corresponding theoretical predictions.  (c, d)  Signal and stokes amplitude spectra at an optical depth of $2d=110$.  (c$'$, d$'$) Corresponding theoretical predictions.}%
\label{figure3}
\end{center}
\end{figure*}
We record the transmission spectra for continuous signal and Stokes fields by sweeping the EOM frequency, which
simultaneously changes the two-photon detuning $\delta$, and measuring their amplitude variations after the
cell. Fig.~\ref{figure2}(a,b) depicts the experimental results for an optical depth of $2d=52$. Solid lines
represent the spectra corresponding to the signal field transmission; dashed lines correspond to the Stokes
field transmission spectra.  Black lines are with no Stokes filtering (i.e., $f=1$); red lines are with a Stokes
intensity attenuation so that $f=\sqrt{0.05}$.  These two values of $f$ ($1$ and $\sqrt{0.05}$) are shown by
horizontal dashed blue lines in Figs.~\ref{figure2}(b,b$'$) and represent the input Stokes amplitude.
With the reduced Stokes seed (red curves), the effects of FWM are suppressed, and we observe a slightly
amplified and nearly symmetric EIT transmission peak.  However, when the full Stokes seed field is present
(black curves), on one hand we observe more gain in the signal field, but on the other hand the FWM/EIT
destructive interference becomes more evident by the presence of a ``knee'' in the signal spectra for small
negative detunings. These results are in very good agreement with the predictions of the theory [see
Fig.~\eqref{figure2}(a$'$, b$'$)], which are calculated from the full expressions in Eqs.~\eqref{signalexpr} and
\eqref{stokesexpr} with no free parameters, where $\Omega$, $\gamma_0$, $d$, and $\gamma$ were computed as in
Ref. \cite{phillipsPRA08}.

Spectra taken at higher optical depth reveal more clear evidence of the constructive and destructive
interference between EIT and FWM. Fig.~\ref{figure2}(c,d) presents similar spectra recorded for the same
signal and Stokes fields, but at an optical depth of $2d=98$. In the case of no Stokes attenuation (black curves), the theoretical curve in Fig.~\ref{figure2}(c$'$) exhibits, as expected, destructive interference at $\dt = \delta - \delta_s= n\pi v_g/L$ for $n = -3, -1$, and $2$ [here $\pi v_g/L =(2 \pi) 31$ kHz and $\delta_s = (2 \pi) 36$ kHz]. While slightly shifted, these three points of destructive interference are also clearly visible in the experimental measurement of Fig.~\ref{figure2}(c).
Even when the Stokes seed field is suppressed, its presence leads to significant distortions in the signal transmission resonance.

We repeated similar spectral measurements after reducing the diameter of the beam by a factor of $1.5$,
which increased the control field Rabi frequency to $\Omega/2\pi = 14$ MHz, and corresponded to a larger light
shift of $\ds/2\pi \approx 85$ kHz. Larger control intensity and smaller beam size allowed us to reduce the control field absorption at high optical
depths and stay within the theoretical model's approximations. Fig.~\ref{figure3} shows the experimental
and corresponding theoretical spectra for optical depths of 52 and 110.  The larger Rabi frequency results in a
larger $v_g$ than above, and thus more closely spaced spectral dips and peaks. There is an excellent agreement
between the experiment and the theory for the Stokes spectra, and for the signal spectra at negative detuning.
However, the theoretical model for signal transmission diverges from experimental observations at positive detuning, indicating the
presence of some unaccounted mechanisms such as nonunity control field refractive index, 
atomic diffusion~\cite{firstenbergPRL09}, and/or the effects of the multi-level structure of the atoms.

\section{Slow Light Measurements}
In this section we discuss the slow light regime for the signal field pulses in the presence of the
co-propagating seeded Stokes field. In particular, we are interested in the prospect of manipulating the signal
pulse group delay and amplitude via the controllable amplitude of the input Stokes field. It is convenient to
use Eqs.~\eqref{signalexpr}, \eqref{stokesexpr} to analyze the dynamics of each pulse propagation through the
medium by calculating the variation in the complex amplitudes of both fields for each spectral component of the
input pulse and then by Fourier transforming the resulting expressions back into the temporal domain.
The group delay of the signal field is determined from the acquired phase, which consists of two contributions. The
first from the first exponential in Eq.\eqref{signalexpr}, and it is the same for all spectral
components of the pulse: $\tau_0 = \frac{d}{d\omega} {\rm Re} \left[ \sigma(\omega) z \right] \approx \frac{d
\gamma z}{2 L \Omega^2}$. Notably, this value is exactly half of the pulse delay expected from the pure
EIT system. The second contribution is from the expression in brackets in Eq.~\eqref{signalexpr}.  The
value of this additional delay depends explicitly on the detuning of the signal pulse from resonance 
and may
vary significantly for different spectral components of the pulse. Below we discussed three distinct scenarios
for the pulse two-photon detuning: $\delta=2|\Delta_{\rm R}|$, $\delta=2\ds$, and $\delta=0$.

\begin{figure*}[hdtp]
\begin{center}
\includegraphics[width=1.6\columnwidth]{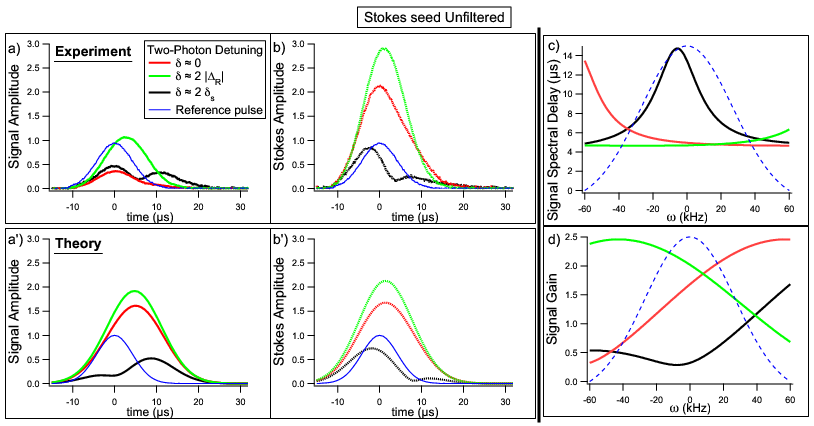}%
\caption{(Color Online)  (a) Slow light on the signal channel at an optical depth of $2d=110$ with no Stokes
seed attenuation (i.e., $f=1$) for signal detunings $\delta = 2|\Delta_{\rm R}|$ (red curves), $\delta = 2\ds$
(green curves), and $\delta = 0$ (black curves).  The thin blue curve is the initial reference pulse. (b)
Corresponding Stokes channel.  (a$'$, b$'$) Corresponding theoretical predictions from
Eqs~\eqref{signalexpr},\eqref{stokesexpr}.  (c)  Theoretical total delay dispersion experienced by the signal
pulse frequency components after traversing length $L$.  The blue dashed curve depicts the frequency spread of
the input signal pulse.  (d)  Predicted signal gain dispersion.} \label{figure4}
\end{center}
\end{figure*}

\begin{figure*}[hdtp]
\begin{center}
\includegraphics[width=1.6\columnwidth]{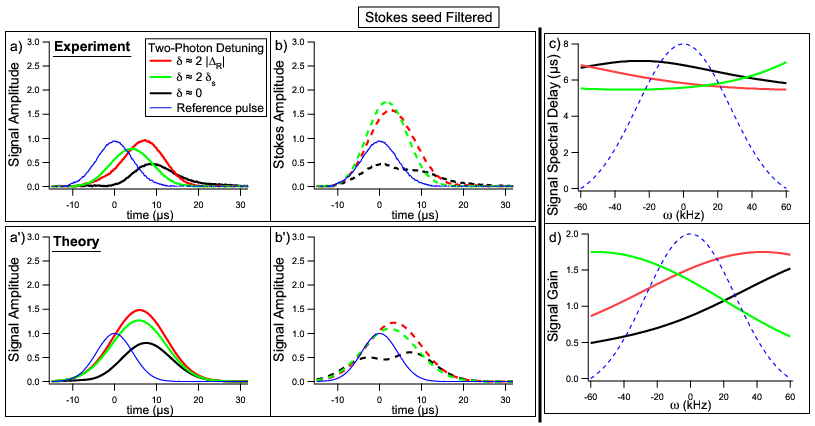}%
\caption{(Color Online)  Same as Fig.~\ref{figure4}, but with Stokes seed attenuation ($f=\sqrt{0.05}$).}
\label{figure5}
\end{center}
\end{figure*}

Figs.~\ref{figure4}(a,b) and \ref{figure5}(a,b) correspondingly present the experimental data for  \mbox{6
$\mu$s}-long (FWHM) signal and Stokes pulses (which corresponds to a bandwidth of $\pm(2 \pi)31$~kHz around the
carrier frequency) when the full Stokes field is present, and when the Stokes field is suppressed. Respective
graphs (a$'$) and (b$'$) give the prediction of the calculations based on the complete solutions of
Eqs.~\eqref{signalexpr},\eqref{stokesexpr}. In these calculations, we use a control field with Rabi frequency
$\Omega/2\pi = 14$ MHz, corresponding to $\delta_s/2\pi =84$ kHz and $\Delta_{\rm R}/2\pi = -28$~kHz.

For more insight into the spectral dynamics of the pulse, we also plot the calculated time delay experienced by the
signal field spectral components $\omega$ [Figs.~\ref{figure4}(c) and ~\ref{figure5}(c)], and the spectral gain
$|\e(\omega,L)|/|\e(\omega,0)|$ [Figs.~\ref{figure4}(d) and ~\ref{figure5}(d)]. These last graphs also show the
spectral bandwidth of the input pulse (the blue, dashed curve) for reference.

\textbf{Case I: $\delta=2|\Delta_{\rm R}|$}

The data corresponding to the case $\delta=2|\Delta_{\rm R}|= (2 \pi) 56$~kHz are shown in Figs.~\ref{figure4}
and \ref{figure5} in red. Although the theoretical calculations predict only a weak dependence of the signal
pulse on the amplitude of the Stokes seed, the experimental data show much stronger dependence: for the full
seeded Stokes field the experimental pulse shows small delay and noticeable attenuation, while when the Stokes
field is small, the signal pulse experiences some small gain and much larger delay. The latter is much closer to
the theoretical expectations of gain of $\approx 1.5$ and the delay of $\approx 6\mu$s. The experimental results
for the Stokes field, however, match the theory much more accurately, which may indicate that the absorption of
the signal field is underrepresented by the model.

The analysis of the spectral gain and delay for both cases ($f = 1$ and $f = \sqrt{0.05}$) provides some
qualitative understanding of the observed pulse behavior. For the case of the full Stokes field,
Fig.~\ref{figure4}(c) shows that all the signal spectral components with $\omega>0$ experience a roughly uniform
delay of $\approx 5\mu$s, whereas the components $\omega<0$ experience a longer delay, resulting in the pulse
spreading as it traverses the cell. However, this effect is somewhat suppressed by lower spectral gain for
$\omega<0$ (Fig.~\ref{figure4}d). With the Stokes field attenuated [Fig.~\ref{figure5}(c,d)], we expect that all
signal spectral components $\omega$ experience a uniform delay of $\approx 6\mu$s, and a slight gain, resulting
in delayed propagation with little pulseshape distortion, as corroborated well in Fig.~\ref{figure5}(a).

\textbf{Case II: $\delta=2\ds$}

The green curves in Figs.~\ref{figure4} and \ref{figure5} depict the results of slow light experiments with a two-photon detuning
of $\delta=2\ds=(2 \pi) 168$ kHz.  Figs.~\ref{figure4}(c,d) illustrate that for the unfiltered Stokes field
($f=1$), all frequency components of the initial signal pulse experience a nearly identical delay of $\approx
5\mu$s---indicating very little pulse spread. Simultaneously, the central component should be amplified by a
factor of $\approx 1.8$. This prediction matches well with both the experimental [Figs.~\ref{figure4}(a,b)] and
theoretical [Figs.~\ref{figure4}(a$'$,b$'$)] pulses.  When the Stokes seed is attenuated ($f=\sqrt{0.05}$), the
signal pulse experiences a slightly longer delay of $\approx5.8\mu$s, but will also be 
less amplified, according to Fig.~\ref{figure5}(d).  The experimental result [green curve in Fig.~\ref{figure5}(a)]
reproduces this predicted delay, but shows a small attenuation rather than gain, possibly indicating the presence of an additional decay mechanism.
%

\textbf{Case III: $\delta=0$}
The black curves in Figs.~\ref{figure4} and \ref{figure5} depict the results of slow light experiments with a
two-photon detuning of $\delta=0$.  This case most clearly demonstrates the merits of Stokes seed attenuation.
For the unfiltered Stokes seed $f=1$, different spectral components will acquire very different phase and gain
while propagating through the interaction region. In particular, Fig~\ref{figure4}(c) shows large variation in
the spectral delay --- from $14\mu$s for central frequencies to $5\mu$s for the farther detuned components. When
combined with the gain curve, shown in Fig.~\ref{figure4}(d), such variations should greatly distort the shape
of the output pulses. In fact, the expected output closely resembles a double-peaked pulse, and is quite similar
to that observed in the experiment. When the Stokes seed is filtered ($f=\sqrt{0.05}$), as in
Fig.~\ref{figure5}, the differential delay is suppressed, and all spectral components experience a common delay
of nearly $7\mu$s, but at the sacrifice of gain, which is $<1$. Figs.~\ref{figure5}(a,b) show excellent
agreement with the corresponding theoretical curves.

\section{Conclusions}
In conclusion, we have demonstrated that both steady-state and dynamic properties of the signal field
propagating under the EIT conditions are strongly effected by resonant four-wave mixing that arise under the
conditions of EIT at high optical depth. This process is well-modeled by a simple double-$\Lambda$ system, where
the output signal and stokes field amplitudes are the results of interference of ``traditional'' EIT and FWM. We
have shown that by attenuating the amplitude of the seeded Stokes field, we can partially control the optical
properties of the medium for the signal field. Moreover, by adjusting the central frequency of the input signal
field around two-photon resonance, in the presence of Stokes seed field we can achieve longer pulse delay and/or
amplification of the signal pulse.

The authors thank M. D. Lukin for useful discussions.  This research was supported by NSF grant PHY-0758010,
Jeffress Research grant J-847, and the College of William~\&~Mary.


\begin{thebibliography}{23}
\bibitem{rabicomment}
As in Ref. \cite{phillipsPRA08}, we define the Rabi frequency $\Omega$ as $|\Omega|^2 = \wp_{es}^2 I/(2 \hbar^2 \varepsilon_0 c)$, where $I$ is the control field intensity.
\end{thebibliography}
\end{document}